\magnification=1200



\def\sqr#1#2{{\vcenter{\hrule height.#2pt
	     \hbox{\vrule width.#2pt height#1pt \kern#1pt
	     \vrule width.#2pt}
	     \hrule height.#2pt}}}
\def\boxx{\hskip1pt\mathchoice\sqr54\sqr54\sqr{3.2}4\sqr{2.6}4
						       \hskip1pt}

\baselineskip=15pt

\def\a{\alpha}      \def\l{\lambda}

                  \def\s{\sigma}  
\def\t{\theta}                        \def\vp{\varphi}
                   \def\o{\omega}  

\def\cO{{\cal O}}

\def\subsection#1{
  \vskip.4cm\goodbreak
  \noindent{\bf #1} 
  \vskip.3cm\nobreak}

\def\mboxx{\buildrel m \over \boxx\,}
\def\mmboxx{\buildrel (m) \over \boxx}
\def\part{\partial}

\font\eightrm=cmr8                                
\font\tfont=cmbx12                                

\vglue 1cm

\centerline{\tfont{Gravitational kinks in two spacetime dimensions}}
\vskip 1cm
\centerline{M Vasili\'c \footnote\dag{\eightrm E--mail: 
	    mvasilic@rt270.vin.bg.ac.yu} 
	   and T Vuka\v sinac \footnote\ddag{\eightrm E--mail: 
	   tatjanav@rt270.vin.bg.ac.yu}}

\centerline{\it Department of theoretical physics, Institute "Vin\v ca",}
\centerline{\it P.O.Box 522, Belgrade, Yugoslavia}
\vskip 2.5cm
\noindent{\bf Abstract.}\hskip .4cm
The properties of gravitational kinks are studied within some
simple models of two dimensional gravity. In spacetimes of cylindrical
topology we prove the existence of kinks of constant curvature with
arbitrary kink numbers. In $R^1\times R^1$ spacetimes $m=1$ kink 
solutions of the equation $R=0$ are found, whereas $|m|>1$ flat kinks
are proved not to exist. We give a detailed analysis of the behaviour of
gravitational kinks under coordinate transformations. Viewed as nonsingular
black holes $|m|>1$ kink solutions are found within a simple dilaton gravity
theory. The general form of the potential function is determined  
from the demand that the theory possesses an arbitrary number of inequivalent 
kink configurations.  
\vskip 1cm
\noindent PACS numbers: 0450, 0240

\vfill\eject

\subsection{1. Introduction}
Over the years there has been a considerable interest in two dimensional
gravity theories. This is primarily because, apart from being
interesting in themselves, these models are simple
enough to serve as useful tools for studying effects of mathematically more
complicated four dimensional spacetime physics.

The two dimensional problem we are concerned with in this paper is related to
the existence of four dimensional gravitational kinks of Finkelstein and
Misner [1]. In a number of papers [2--9] the properties of these topologically 
nontrivial metric configurations have been analysed. They have been shown
to possess a {\it twisting light--cone structure} with regions of anomalous 
causality. Each twist, being surrounded by a one--way surface through which 
causes can propagate in only one direction, is an example of a nonsingular
black hole. A dynamical model containing solutions of this type is,
however, missing. It has further been shown that the configuration space of
four dimensional kinks is double connected providing a basis for the
existence of double--valued wave functions. Unfortunately, the hope that
gravitational kinks could be quantized as fermions has disappeared after it 
has been realized that each homotopy class of metric configurations contains
a spherically symmetric one [10]. Spherical symmetry of four dimensional
kinks tells us that they are basically two dimensional objects. The
mechanism for their existence is a feature of two dimensional
spacetime. Indeed, it has been shown in reference [11] that two dimensional
metrics with Minkowski signature possess similar homotopy structure to that
of four dimensional case.

The purpose of this paper is to study the properties of two dimensional
gravitational kinks within some simple physical models. In section two the
homotopy structure of the space of metric components is analysed and the
generalized Liouville equation in spacetimes of $R^1\times S^1$ topology
is solved showing that gravitational kinks of constant curvature
do exist on a cylinder. These solutions are, however, inappropriate to be
viewed as nonsingular black holes since their horizons are uniformly 
disposed over the space. Section three is devoted to the analysis of 
kink--type nonsingular black holes in spacetimes
of trivial topology. The equation $R=0$ is solved using generalized light 
cone coordinates. Only one--kink regular solutions are found to exist in
accordance with the simple observation that more than one kink is needed
to form a genuine black hole. The transformation properties of gravitational
kinks are analysed in detail. In section four a model of two dimensional
gravity coupled to a nonphysical dilaton field is considered. The potential
term is determined from the requirement that the theory admits an arbitrary
number of inequivalent stationary kink configurations. 
\subsection{2. Kinks on a cylinder}
In two spacetime dimensions the space of components of the metric tensor
$g_{\mu\nu}$ is three dimensional. There is, however, a forbidden region in
this space defined by vanishing metric determinant. The corresponding
equation
$ g\equiv g_{00}g_{11}-(g_{01})^2 = 0 \  $
defines a cone--shaped two dimensional surface which devides the space of
metric components into three separate regions each characterized by a
definite signature. The figure 1 clearly shows that the only topologically
nontrivial region is that of Minkowski signature. It is obviously 
multiply connected with the first homotopy group being the group of integers.

In spacetimes of $R^1\times S^1$ topology the corresponding configuration
space is represented by closed loops which completely lie in one connected
region. A noncontractible loop, $m$ times wound around the forbidden
surface $g=0$, is called a kink and is attributed the kink number $m$.
Let us choose representatives, one in each homotopy class of metric
configurations, to be
$$ g_{\mu\nu}^m \equiv \pmatrix{ \cos{m\s}&\sin{m\s}\cr
				 \sin{m\s}&-\cos{m\s}\cr}     \eqno(1)$$
where $\s$ runs from 0 to $2\pi$ and $m$ is an integer. The metric 
(1) is represented by a loop which leaves the point $g_{\mu\nu}
=\eta_{\mu\nu}$ (denoted by $M$ in figure 1), 
winds $m$ times around the surface
$g=0$ and ends in $g_{\mu\nu}=\eta_{\mu\nu}$ again. It is obvious that no
continuous coordinate transfomation can map kinks with different kink
numbers into each other. The discrete transformation of time reversal,
however, maps kinks into antikinks [1].
The light cone structure of the $g_{\mu\nu}^m$ kinks is found to be defined
by the equation
$$ {d\tau\over d\s} = {\cos{m\s}\over \sin{m\s} \pm 1} \ \, .  \eqno(2)$$
As shown in figure 2 the light cones twist along the $\s$--axis just as light 
cones of Finkelstein and Misner do. The $m$--kink is characterised by $m/2$
twists.

Let us now try to find a physical model possessing kink solutions on a 
cylinder. The simplest equation $R=0$ is a two dimensional analogue of the
free Einstein equation in four dimensions. Can gravitational kinks on
a cylinder be flat?
As shown in reference [11] any two dimensional metric with Minkowski
signature can be written in the form 
$$g_{\mu\nu} = e^{\vp} \pmatrix{ \cos\psi & \sin\psi \cr
				 \sin\psi &-\cos\psi\cr} \ \, .  $$
One of the functions $\vp$ or $\psi$ could further be specified but the
usual choice $\psi =0$ is too restrictive excluding all kink metrics. Our
choice of gauge,
$$ g_{\mu\nu} = e^\vp g_{\mu\nu}^m                               $$
is quite acceptable in that respect, although also incomplete.
With this gauge choice the equation $R=0$ reduces to
$$ {\mboxx}\vp + m^2\cos{m\s} = 0                         \eqno(3)$$
where $\mboxx$ stands for the covariant D'Alambertian of the metric 
$g_{\mu\nu}^m$. To solve the equation (3) we shall make use of the 
generalized light cone coordinates
$$x^\pm \equiv {1\over \sqrt 2} [\,\tau - {1\over m} \ln{(1\mp\sin{m\s})}\,]
								   $$
obtained by integrating the equations (2). The new coordinates are well
defined only in regions of nonvanishing $\,\cos{m\s}$.
Skipping the details of the calculation we just present the result. The
general solution of the equation (3) is given by
$$ \vp = -m\tau + A(x^+) + B(x^-)                                 $$
where $A$ and $B$ are two arbitrary functions. Since our light cone 
coordinates are defined only between zeros of $\,\cos{m\s}\,$ we must
separately check the regularity of $\vp$. A simple observation 
$${\bar x}^\pm\equiv e^{-m\sqrt 2 x^\pm}=(1\mp\sin{m\s}) e^{-m\tau} $$
leads us directly to the manifestly regular form of $\vp$,
$$\vp = -m\tau + {\bar A}({\bar x}^+) + {\bar B}({\bar x}^-) \ \, . $$
Therefore, flat kinks do exist on a cylinder. Note, that this conclusion
is not in discrepancy with the known result that it is always possible
to find coordinate transformations which map an arbitrary flat metric into
the Minkowski metric. Namely, this theorem is valid only locally and only in
simply connected regions of a manifold. This is certainly not the case 
with the regions which accommodate kinks on a cylinder.

Let us briefly mention in the end of this section that kinks of constant 
curvature also exist on a cylinder. The equation $R=\l$ leads to the 
generalized Liouville equation
$${\mboxx}\vp - \l\, e^\vp + m^2\cos{m\s} = 0                       $$
with the general solution of the form
$$\l\, e^\vp = {4\over \cos{m\s}}\, {(\part_+A)(\part_-B)\over (1-AB)^2} 
						     \ \, .        $$
A tedious analysis shows that the functions $A(x^+)$ and $B(x^-)$ can
be chosen to ensure regularity of the solutions.
\subsection{3. Kinks in $R^1\times R^1$ spacetimes}
Gravitational kinks on a cylinder are not good analogues of realistic
nonsingular black holes because their horizons are disposed all over
the space. For that reason,
in this section, we analyse 2d gravitational kinks in spacetimes of trivial
topology. Homotopy considerations are the same as in the preceding section
provided the boundary values of the metric are kept fixed.
In each homotopy class we choose an asymptotically Minkowskian representative
$$g_{\mu\nu}^{(m)} \equiv \pmatrix{ \cos{m\t} & \sin{m\t}\cr
			   \sin{m\t} &-\cos{m\t}\cr}                $$
with $\t$ given by 
$\t \equiv \pi + 2\,{\rm arctg}\,\s\    $
and $\s$ running from $-\infty$ to $+\infty$. It has been suggested [11]
that one could make use of a singular coordinate transformation to obtain
$R^1\times R^1$ kinks from $R^1\times S^1$ ones. We do not use this
technique because it leads to geodesically incomplete solutions even
in the asymptotic region which is highly undesirable if we want to view
gravitational kinks as analogues of realistic black holes.
Instead, we restrict ourselves to the metric configurations of the form
$$ g_{\mu\nu} = e^\vp g_{\mu\nu}^{(m)} \ \, .                \eqno(4)$$
The light cone structure of this geometry has the twisting nature of figure 2
but is asymptotically Minkowskian. The lines $\,\cos{m\t}=0\,$ define the
horizons of the corresponding black hole.

Now, let us try to solve the equation $R=0$ in the gauge (4). First, we
solve the equation $ds^2=0$ and define the light cone coordinates
$$ y^{\pm}\equiv{1\over\sqrt 2}[\,\tau - s_m(\mp\t )\,]       \eqno(5)$$
with
$$ s_m(\t )\equiv \int {\cos{m\t}\, d\t\over (1-\cos{\t})(1+\sin{m\t})}\ \, .
							       \eqno(6)$$
The coordinates $y^\pm$ are well defined only between zeros of the function
$\,\cos{m\t}\,$. The equation $R=0$, rewritten as
$$ {\mmboxx}\vp - (\cos{m\t})'' = 0                           \eqno(7)$$
(primes denote differentiation over $\s$) has a particular solution of the 
form  
$$ \vp_{par}=-\ln{|\cos{m\t}|}\ \, .                          \eqno(8)$$
On the other hand, the homogeneous equation is most easily solved in the
light--cone coordinates (5) in which
$$ {\mmboxx}\vp = {2\over \cos{m\t}}\, \part_+\part_-\vp \ \, .      $$
Using this fact we find the general solution of the equation (7) to be
$$ \vp = A(y^+) + B(y^-) - \ln{|\cos{m\t}|} \ \, .            \eqno(9)$$
This solution is manifestly regular only in regions of nonvanishing
$\,\cos{m\t}\,$.
Is it possible to choose the functions $A$ and $B$ and make $\vp$ regular 
everywhere? 

In the simplest case $m=1$ one easily solves the integral (6) and finds
$$ y^\pm = {1\over\sqrt 2}(\,\tau\pm\s - 2\ln{|\s\pm 1|}\,)          $$
whereas the particular solution (8) boils down to
$$ \vp_{par}={1\over\sqrt 2} (y^+ +y^-) + \ln{(1+\s^2)} - \tau \ \, . $$
Consequently, the general solution (9) can be rewritten in the form
$$ \vp = C(y^+) + D(y^-) + \ln{(1+\s^2)} - \tau              \eqno(10)$$
or, if we use the coordinates
$$ {\bar y}^\pm \equiv e^{-\sqrt 2 y^\pm} = (1\pm\s )^2 e^{-\tau\mp\s}   $$
in the manifestly regular form
$$\vp ={\bar C}({\bar y}^+)+{\bar D}({\bar y}^-)+\ln{(1+\s^2)}-\tau\ \, .  $$
By choosing the functions ${\bar C}$ and ${\bar D}$ to be everywhere well 
defined we prove the existence of flat $m=1$ kinks in spacetimes of trivial
topology. An example of this kind is considered in reference [11].

Although kinks cannot be deformed into no--kink configurations the fact
that $R=0$ tells us that there must exist a continuous transformation
which maps between (10) and the Minkowski solution.  
It is not difficult to explain the apparent inconsistency of this result.
As we mentioned at the beginning
of this section the homotopy considerations in topologically trivial
spacetimes make sense only if the boundary values of the metric are kept fixed.
The fact that a one--kink solution can be transformed into a no--kink solution
only means that the corresponding coordinate transformation does change
the asymptotics. To see this we shall make use of figure 2 where twisting
light cones of an $m=1$ kink are displayed, and try to draw purely spacelike
and timelike coordinate lines. These are necessary ingredients of any 
coordinate system in which light cones do not twist. The result is shown
in figure 3. We see that it is indeed possible to construct no--kink 
coordinates but also that it is impossible to do so without moving the
points of $\s$--infinity.
The analysis of figure 3 is in agreement with Finkelstein's argument that
spacetimes which admit a global family of spacelike surfaces do not have
actual kinks [10].

Similar analysis can also be done for $|m|>1$. The set of timelike and spacelike
lines for a two--kink metric is shown in figure 4. We see that these
lines cannot be used as coordinate lines of a global coordinate system. 
Consequently, it is not possible to transform
$m=2$ kinks (and the same holds for $|m|>2$) into
no--kink configurations. We conclude that regular $|m|>1$ solutions of the
equation $R=0$ do not exist.
To support this conclusion we shall analyse the behaviour of the general
solution (9) in the neighbourhood of the light cone singular points. 
Without loss
of generality we restrict ourselves to the case $m=2$. It is easy, then, to
show that there are four singular points --- horizons,
$$\s_1=-(1+\sqrt 2)\ \,  \qquad \s_2=1-\sqrt 2\ \, $$
$$\s_3=-(1-\sqrt 2)\ \,  \qquad \s_4=1+\sqrt 2\ \, $$
which devide the $\s$--axis into five separate regions. In each region we are
free to choose the functions $A$ and $B$ independently. To be able to 
analyse their behaviour in the vicinity of the horizons we first solve the
integral (6) and find the explicit form of the light cone coordinates:
$$\eqalign{
&y^+ \sim \s_1\ln{|\s -\s_1|} - \s_3\ln{|\s -\s_3|} \ \, \cr
&y^- \sim \s_2\ln{|\s -\s_2|} -\s_4\ln{|\s -\s_4|} \ \, .}            $$
At the same time the behaviour of the particular solution (8) is given by
$$\ln{|\cos{2\t}|}\sim \ln{|\s -\s_1|} + \ln{|\s -\s_2|}
      +\ln{|\s - \s_3|} + \ln{|\s -\s_4|} \ \, .             \eqno(11)$$
We see that the functions $A$ and $B$ must diverge linearly to cancel the
infinite contributions of (11). Let us consider the following two regions
of the $\s$--axis: $(\s_1,\s_2)$ and $(\s_2,\s_3)$. The corresponding 
$A$ functions we shall denote by $A_1$ and $A_2$. Then, the requirement
that the function $\vp$ be finite at the points $\s =\s_1$ and $\s =\s_3$
gives 
$$A_1\sim {1\over\s_1}\, y^+\ \,  \ \ {\rm and}\ \  
  A_2\sim -{1\over\s_3}\, y^+ \ \, .                                 $$
We notice that the rates of decrease of the two functions differ. On the
other hand, the demand for continuity of the function $\vp$ at the point
$\s = \s_2$, where $y^+$ is finite, gives
$$A_1(y^+) = A_2(y^+) + {\rm const}\ \, .                            $$
This constraint implies identical asymptotic behaviour of the functions
$A_1$ and $A_2$, which contradicts their formerly established asymptotics.
We conclude that regular two--kink solutions of the equation $R=0$ in
topologically trivial spacetimes do not exist. The same holds for $|m|>2$.

For the analysis we just presented the gauge choice (4) was not essential.
There are metric configurations, not covered by our gauge choice, which
also have twisting light cones even if their kink number is zero. An
example is the loop of figure 1 $n$ times wound around the forbidden surface
and then $n$ times unwound in the opposite direction. Such configurations do
not admit global families of spacelike lines and, consequently, cannot be
flat. They are examples of $m=0$ nonsingular black holes.
\subsection{4. Kinks in dilatonic gravity}
In quest of a physical model accommodating gravitational kinks as analogues
of four dimensional nonsingular black holes we shall consider a simple
dilaton gravity theory given by the action
$$I=\int d^2x\, \sqrt{-g}\,[\,\phi R + V(\phi )\,] \ \, .   \eqno(12)$$
This kind of theory has already been considered in reference [12] where the
form of the potential has been determined from the requirement that the
theory admits nonsingular black hole solutions. Similar constructions of
nonsingular black holes can also be found in references [13-18]. All these,
however, are not kink type solutions. Following the idea of [12],
recently developed in [19], we search for the potential $V(\phi )$ which
admits stationary kink configurations. Variation of the action (12) yields
the field equations
$$ R+{dV\over d\phi}=0                                      \eqno(13)$$
and
$$\nabla_\mu\nabla_\nu\phi + {1\over 2}\, g_{\mu\nu}V(\phi )=0 
							    \eqno(14)$$
only two of which are independent. As in the preceding section we choose
the gauge (4). Then, the most general stationary solution is obtained by
substituting $\phi =\phi (\s )$ and $\vp = \vp (\s )$ into the field
equations. Combining the time--time and the space--space components of the
equation (14) we find
$$\phi ''-\vp '\phi ' = 0                                           $$
wherefrom either $\phi = {\rm const.}$ or
$$ e^\vp = \a\phi ' \ \ \ \ \a ={\rm const} \ \, .        \eqno(15)$$
The value of the potential $V(\phi )$ can also be explicitely expressed in
terms of the dilaton field $\phi$ by combining the time--time and the
space--time components of (14). The result is 
$$(\phi '\cos{m\t})' = e^\vp V(\phi ) \ \, .                  \eqno(16)$$

Let us first consider the simplest case $\phi = {\rm const.}$ Then, the
full set of field equations is equivalent to
$$\phi =\phi_0\ \  \qquad V(\phi_0) = 0\ \  \qquad 
R=-\biggl(\, {dV\over d\phi}\, \biggr)_0 \equiv \l           \eqno(17)$$
where $\phi_0$ and $\l$ are constants. If $\l =0\,$, we have already seen
that kink solutions with $|m|>1$ do not exist. If $\l\ne 0$, the stationary
equation $R=\l$ boils down to
$$ \bigl[\, {1\over\o}(\o\cos{m\t})'\, \bigr] ' = \o         \eqno(18)$$
where $\o\equiv \l e^\vp$ and, consequently, $\o\ne 0$ everywhere. When
$|m|>1$ the function $\o\cos{m\t}$ changes sign at least four times. Then, 
$(\o\cos{m\t})'$ changes sign at least three times and so does $(\o\cos{m\t})'
/\o$. It follows then that its derivative must change sign at least twice
which is, according to (18), in contradiction with $\o\ne 0$ everywhere.
Thus, there are no $|m|>1$ kink solutions of the equations (17).

The case $\phi\ne{\rm const.}$ is easily shown to be completely described
by the equations (15) and (16) alone. To simplify further analysis we shall
make use of the prepotential function $W(\phi )$ defined as
$$\a V(\phi ) \equiv {dW(\phi )\over d\phi} \ \, .       \eqno(19)$$
The equation (16) is then easily integrated out to give
$$\phi '\cos{m\t} = W(\phi ) \ \, .                      \eqno(20)$$
Without loss of generality we set the integration constant to zero. 
According to (15) the dilaton field $\phi (\s )$ is a monotonous 
function of $\s$ taking values in the interval $D\equiv (\phi_-,\phi_+)$
defined by the endpoints $\phi (-\infty )$ and $\phi (+\infty )$.
If the interval $D$ is finite we see that $W(\phi )$ must tend to zero
as $\phi$ approaches its boundary. If $D$ is infinite the function $W(\phi )$
can go to infinity but not faster than $\cO (\phi )$. It follows also
that the prepotential function $W(\phi )$ must have the same number of
zeros on $D$ as $\,\cos{m\t}\,$ does on the whole $\s$--axis.
Let us denote these zeros by $\phi_i$ and $\s_i$ respectively ($i=1,...,2m$).
Taking the derivative of (20) then gives another constraint on $W(\phi )$:
$$\biggl(\, {dW\over d\phi}\,\biggr)_{\phi =\phi_i} = 
	\biggl(\, {d\, \cos{m\t}\over d\s}\,\biggr)_{\s =\s_i}
							    \eqno(21a)$$   
where $\phi_i<\phi_{i+1}$ and $\s_i<\s_{i+1}$. Together with
$$\eqalign{
&W(\phi_i) = 0  \cr
&W(\phi )\to 0 \qquad \qquad {\rm as}\qquad\ \phi\to \phi_\mp = 
		{\rm finite} \cr
&{W(\phi )\over \phi}\to{\rm finite}\qquad {\rm as}\qquad\ 
		 \phi\to\phi_\mp = \mp\infty }            \eqno(21b)$$
these constraints represent necessary conditions for the equation (20) to
have well defined solutions. Given a continuous prepotential function 
$W(\phi )$ subject to the above constraints the equation (20) is easily
integrated in each region $\s_i<\s <\s_{i+1}\,$,
$$\int {d\phi\over W(\phi )} = \int {d\s\over \cos{m\t}} + {\rm const}
							   \eqno(22)$$
yielding a continuous and monotonous function $\phi(\s )$ on the whole
$\s$--axis. Furthermore, the integration constants in (22) can be chosen
to ensure its differentiability. If the function $W(\phi )$ has well
defined higher derivatives then $\phi (\s )$ will be of the same kind.
Therefore, {\it the conditions (21) are also sufficient conditions for the field
equations to admit regular m--kink solutions}.

Let us consider some examples. The easiest way to construct a potential
with given properties is to choose a monotonous function $\phi (\s )$,
substitute it into (20) to obtain $W(\phi (\s ))$ as a function of $\s$,
and then use $\s =\s (\phi )$ to find $W(\phi )$ itself. Taking the linear
dilaton vacuum $\phi =\s$ as a starting point we thus find
$$W(\phi )=W_m(\phi )\equiv \cos{\bigl[ m\t (\phi)\bigr]} \ \, .   \eqno(23)$$
In particular, the one--kink and two--kink prepotentials are given by
$$W_1(\phi ) =1-{2\over 1+\phi^2}\ \  \qquad 
  W_2(\phi ) =1-{8\phi^2\over (1+\phi^2)^2} \ \, .                  $$
The potential $V(\phi )$ is determined from (19) up to a multiplicative
constant.

As our second example we shall consider the dilaton field of the form
$\phi = \t (\s )$. It follows then that
$$W(\phi )= (1-\cos{\phi})\cos{m\phi}   \ \, .                 \eqno(24)$$
In its domain $D=(0,2\pi )$ this function has $2m$ zeros and consequently
admits $m$--kink solutions. Notice, however, that, restricted to the
domain between two neighbouring zeros, like, for example,
$D_0\equiv (\pi /2m ,3\pi /2m)$, the prepotential function
$W(\phi )$ has no zeros at all and the equation (22) has regular $m=0$
solutions. If we take $D_1\equiv (\pi -3\pi /2m ,\pi +3\pi /2m)$
as the domain of our function $W(\phi )$ we see that the conditions (21)
for the existence of $m=1$ kinks are fulfilled up to a multiplicative
constant. Since the potential function $V(\phi )$ {\it is} determined
only up to a multiplicative constant, we conclude that the field equations 
possess one--kink solutions. Thus, the choice (24) of the prepotential 
$W(\phi )$ ensures the existence of three inequivalent kink configurations.

In the end, let us comment on the possibility to have a potential which
would allow for an arbitrary number of inequivalent kink configurations.
If $D_m$ ($m=0,1,...,k$) are disjoint finite domains of the prepotential
functions $W_m(\phi )$ then it is possible to construct the function 
$W(\phi )$ with the domain 
$D\supset\bigcup_m D_m$ such that its restriction
to $D_m$ is exactly $W_m$. If $W_m$ admits $m$--kink solutions it follows
that $W$ will admit $k+1$ inequivalent kink configurations with kink
numbers taking values from $0$ to $k$. The theory of this kind will possess
$k-1$ types of nonsingular black holes. Our examples (23) and (24)
describe nonsingular black holes in different background geometries.
The first is asymptotically flat spacetime while the second is a spacetime
of constant negative curvature.
\subsection{5. Concluding remarks}
We have studied in this paper the properties of two dimensional gravitational
kinks as solutions of some simple physical models. In section 2 we have proved
that spacetimes of $R^1\times S^1$ topology admit kink metric configurations
of constant curvature. In particular, flat gravitational kinks of
arbitrary kink number exist on a cylinder. Viewed as nonsingular black
holes, however, these solutions are not appropriate since the horizons
are uniformly disposed over the space. To study more realistic black
hole solutions we switched our attention in section 3 to spacetimes of
trivial topology. We have proved the existence of $m=1$ flat gravitational
kinks and argued that these cannot be black hole solutions since it is
possible to transform them away. The corresponding coordinate transformations 
necessarily move the points of spatial infinity. The formal $|m|>1$ kink
solutions of the equation $R=0$, valid in regions between the horizons, 
have been shown to inevitably contain singularities. It has been explained
in detail why it is impossible to transform $|m|>1$ kinks away. Finally,
a simple model of dilaton gravity theory has been considered. It has been
shown that a nontrivial potential function is needed if one wants the
theory to accommodate nonsingular black hole solutions. The general form
of the potential has been found to ensure the existence of stationary
gravitational kinks and some examples have been given. We have also proved
the possibility of choosing the potential function which admits an
arbitrary number of inequivalent kink configurations.
\subsection{Acknowledgements}
We wish to thank R. Panajotovi\'c for her help in making the illustrations.
This work has been supported in part by the Serbian Research Foundation,
Yugoslavia.

\vfill
\eject
\subsection{References}
\item{[1]} Finkelstein D and Misner C W 1959 
	   {\it Ann. Phys., NY} {\bf 6} 230
\item{[2]} Williams J G and Zia R K P 1973 {\it J. Phys.} A {\bf 6} 1
\item{[3]} Finkelstein D and McCollum G 1975 {\it J. Math. Phys.} {\bf 16} 2250
\item{[4]} Whiston G H 1981 {\it J. Phys.} A {\bf 14} 2861
\item{[5]} Cl\'ement G 1984 {\it Gen. Rel. Grav.} {\bf 16} 131, 477, 491;
	   1986 {\bf 18} 137
\item{[6]} Harriott T A and Williams J G 1986 {\it J. Math. Phys.} 
	   {\bf 27} 2706 
\item{[7]} Dunn K A and Williams J G 1989 {\it J. Math. Phys.} {\bf 30} 87
\item{[8]} Dunn K A 1990 {\it Gen. Rel. Grav.} {\bf 22} 507
\item{[9]} Finkelstein D 1991 Prehistory of the black hole {\it Georgia
	   Institute of Technology} \hfill
\item{}     {\it preprint}
\item{[10]} Finkelstein D 1966 {\it J. Math. Phys.} {\bf 7} 1218
\item{[11]} Dunn K A, Harriott T A and Williams J G 1992
	    {\it J. Math. Phys.} {\bf 33} 1437
\item{[12]} Trodden M, Mukhanov V F and Brandenberger R H 1993
	    {\it Phys. Lett.} {\bf 316B} 483
\item{[13]} Altshuler B 1990 {\it Class. Quant. Grav.} {\bf 7} 189
\item{[14]} Morgan D 1991 {\it Phys. Rev.} D {\bf 43} 3144
\item{[15]} Dymnikova J 1992 {\it Gen. Rel. Grav.} {\bf 24} 235
\item{[16]} Mukhanov V and Brandenberger R 1992 {\it Phys. Rev. Lett.}
	    {\bf 68} 1969
\item{[17]} Banks T and O'Loughlin M 1992 {\it Rutgers preprint} RU-92-61
\item{[18]} Brandenberger R, Mukhanov V and Sornborger A 1993
	    {\it Phys. Rev.} D {\bf 48} 1629
\item{[19]} Chan K C K and Mann R B 1995 {\it Class. Quant. Grav.} 
	    {\bf 12} 1609                    
\eject
\subsection{Figure captions}
\vskip .5cm
{\bf Figure 1.} The surface $g=0$ in the space of metric components

{\bf Figure 2.} One--kink and two--kink light cone configurations

{\bf Figure 3.} Spacelike and timelike coordinate lines for a one--kink metric

{\bf Figure 4.} Spacelike and timelike lines of a two--kink metric
\vfill\break
\eject

\end